\documentclass[12pt,preprint]{aastex}
\newcommand{\pref}{\protect\ref}
\newcommand{\solrad}{\ifmmode{R}_{\rm S}\else${R}_{\rm S}$\fi}
\newcommand{\solmas}{\ifmmode{M}_{\rm S}\else${M}_{\rm S}$\fi}

\newcommand{\ctn}{\ifmmode\kappa\else$\kappa$\fi}

\newcommand{\flxu}{$\,$ergs$\,$cm$^{-2}\,$s$^{-1}$}

\newcommand{\dynu}{$\,$dyn$\,$cm$^{-2}$}
\newcommand{\term}[2]{\mbox{$\,^{#1}{\rm #2}$}}
\def\term#1 #2/{\mbox{$\,^{#1}{\rm #2}$}}

\def\aspcs{{ASP Conf.\ Ser.}} 

\newcommand{\pder}[2]{{{\partial {#1} \over {\partial {#2}}}}}

\def\mathstacksym#1#2#3#4#5{\def#1{\mathrel{\hbox to 0pt{\lower 
    #5\hbox{#3}\hss} \raise #4\hbox{#2}}}}

\mathstacksym\lta{$<$}{$\sim$}{1.5pt}{3.5pt} 
\mathstacksym\gta{$>$}{$\sim$}{1.5pt}{3.5pt} 
\mathstacksym\lrarrow{$\leftarrow$}{$\rightarrow$}{2pt}{1pt} 
\mathstacksym\lessgreat{$>$}{$<$}{3pt}{3pt} 

\renewcommand{\vec}[1]{{\bf #1}}
\newcommand{\cross}{\times}

\newcommand{\jcb}{\ifmmode\vec{j}\cross\vec{B}\else$\vec{j}\cross\vec{B}$ \fi}

\newcommand\figone{
\begin{figure}[!ht] 
\epsscale{0.9}
\plotone{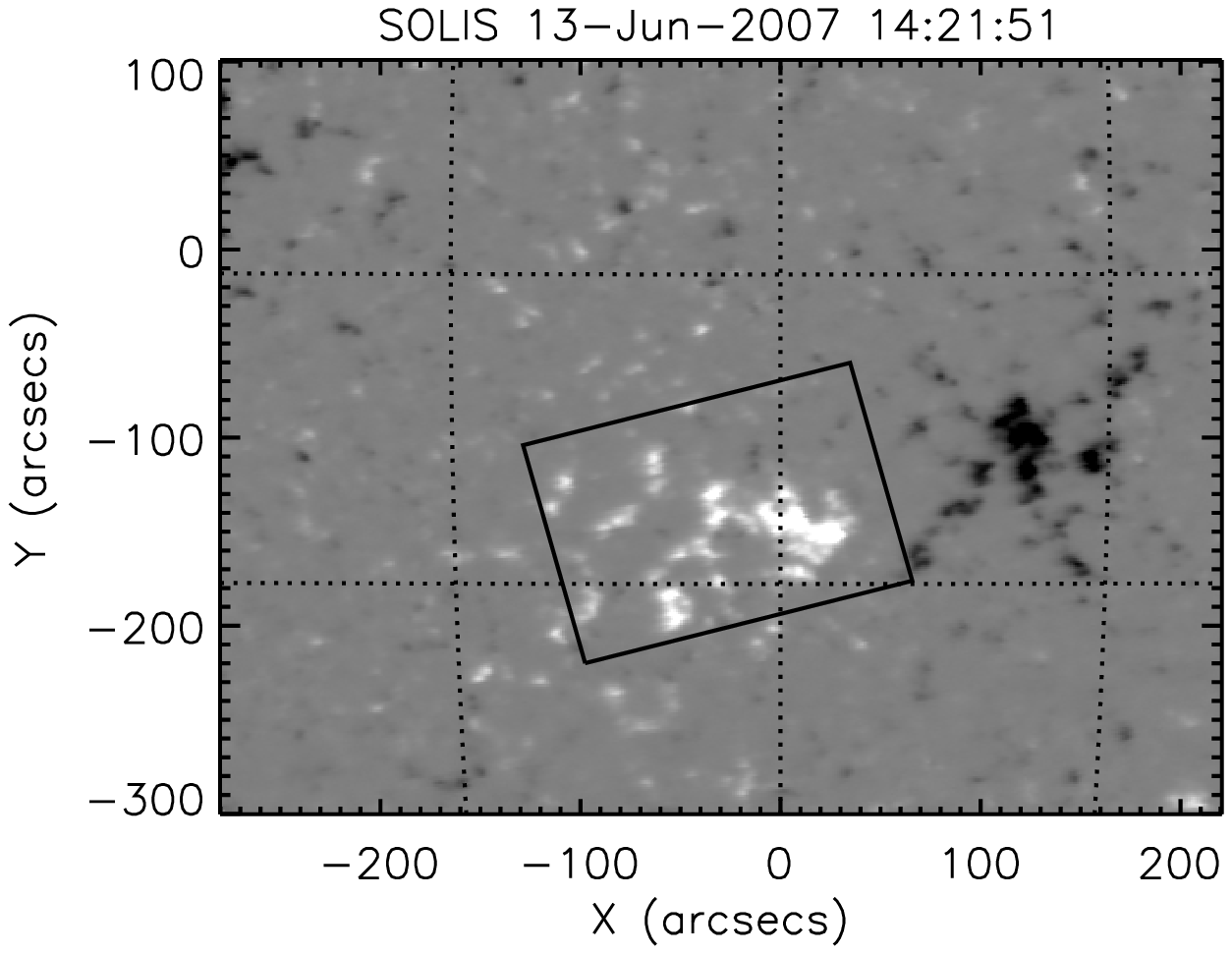}  
\caption{\label{fig:context} A context image of the region observed 
on June 13 2007 by the ESG.  The data show the SOLIS magnetogram 
obtained at 14:21 UT, rotated back to the epoch of the ESG
raster scan (10:52-10:56 UT).  The color table is linear between 
$\pm 150$ Mx~cm$^{-2}$.  The black rectangle indicates the field of view
of the Echelle Spectrograph scan.
}
\end{figure}
}
\newcommand\figtwo{
\begin{figure}[!ht] 
\epsscale{1.1}
\plotone{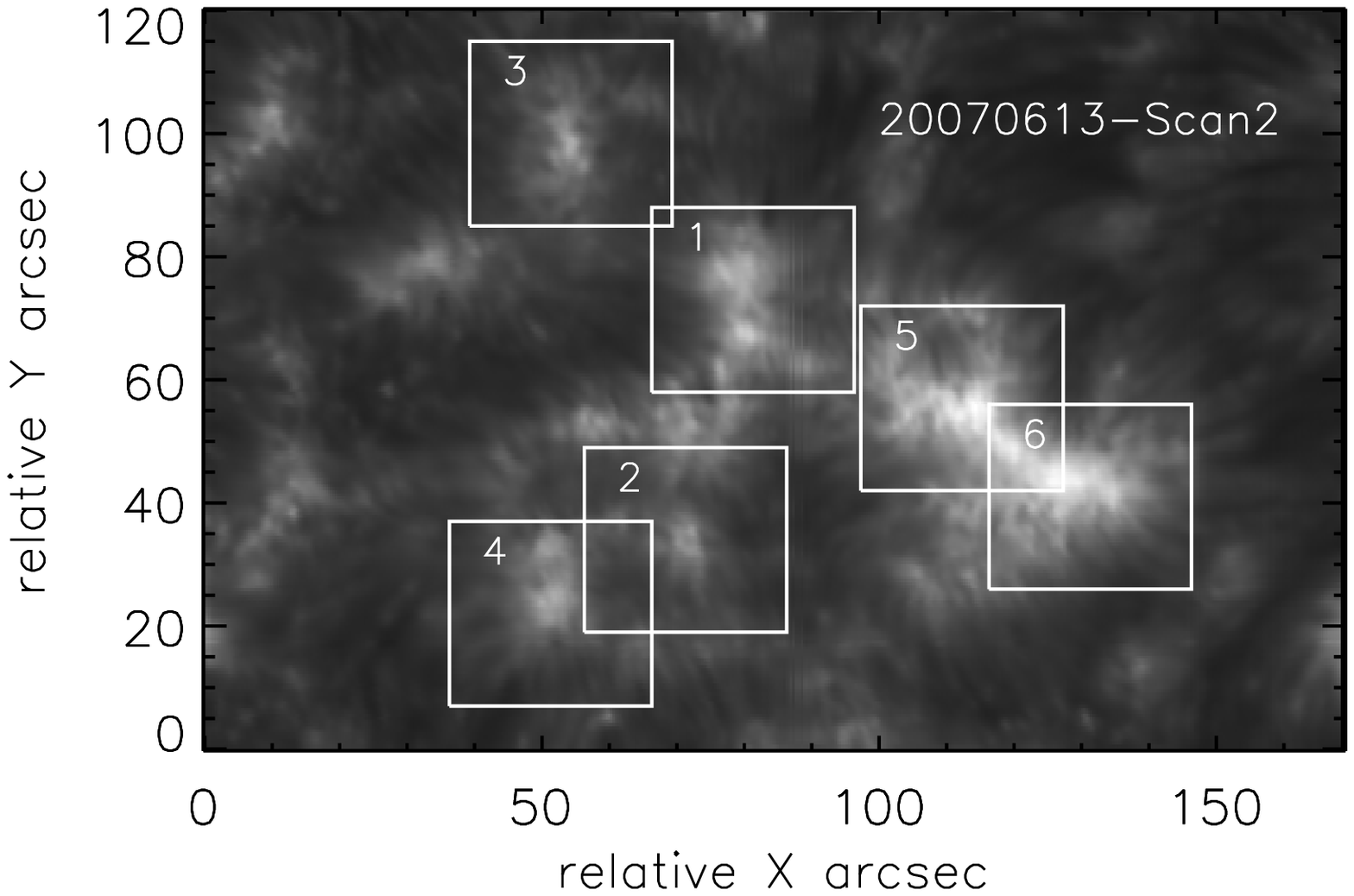}  
\caption{\label{fig:core} The core region of the Ca II H line observed 
on June 13 2007 by the ESG.  The intensities within $\pm 2.3$ m\AA{} of the
line core are 
summed and shown in this plot. The boxes labeled 1 through 6 show
regions studied in detail in later figures and the text.
}
\end{figure}
}

\newcommand\figthree{
\begin{figure}[!ht] 
\epsscale{1.05}
\plotone{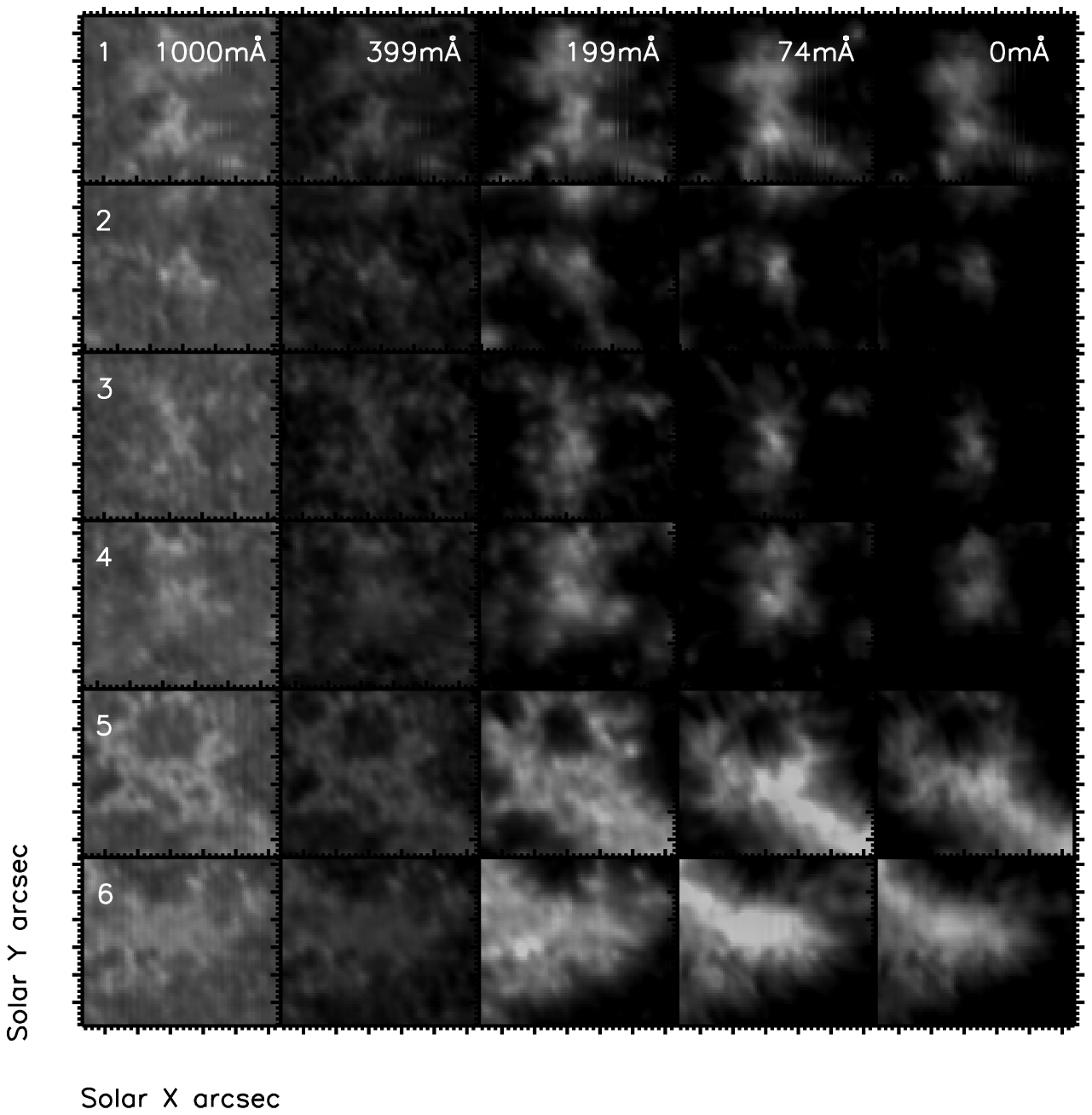}  
\caption{\label{fig:images} ``Monochromatic'' ESG images, binned over
  22.5 m\AA{} are shown 
for each boxed region of figure~\pref{fig:core},
for the data
obtained on 13 June 2007.
  Relative intensities
can be compared between all images.  Each region is
$30\arcsec\times30\arcsec$ in size, slightly smaller than a typical
supergranule cell size.  One minor tick mark corresponds to 1$\arcsec$.  
0 mA corresponds to line center, all other wavelengths are to the
red side.  The color table is linear, ranges between 1/4 and 1 of a
fixed number of counts, and is the same for all frames.
}
\end{figure}
}

\newcommand\figfour{
\begin{figure}[!ht] 
\epsscale{0.6}
\plotone{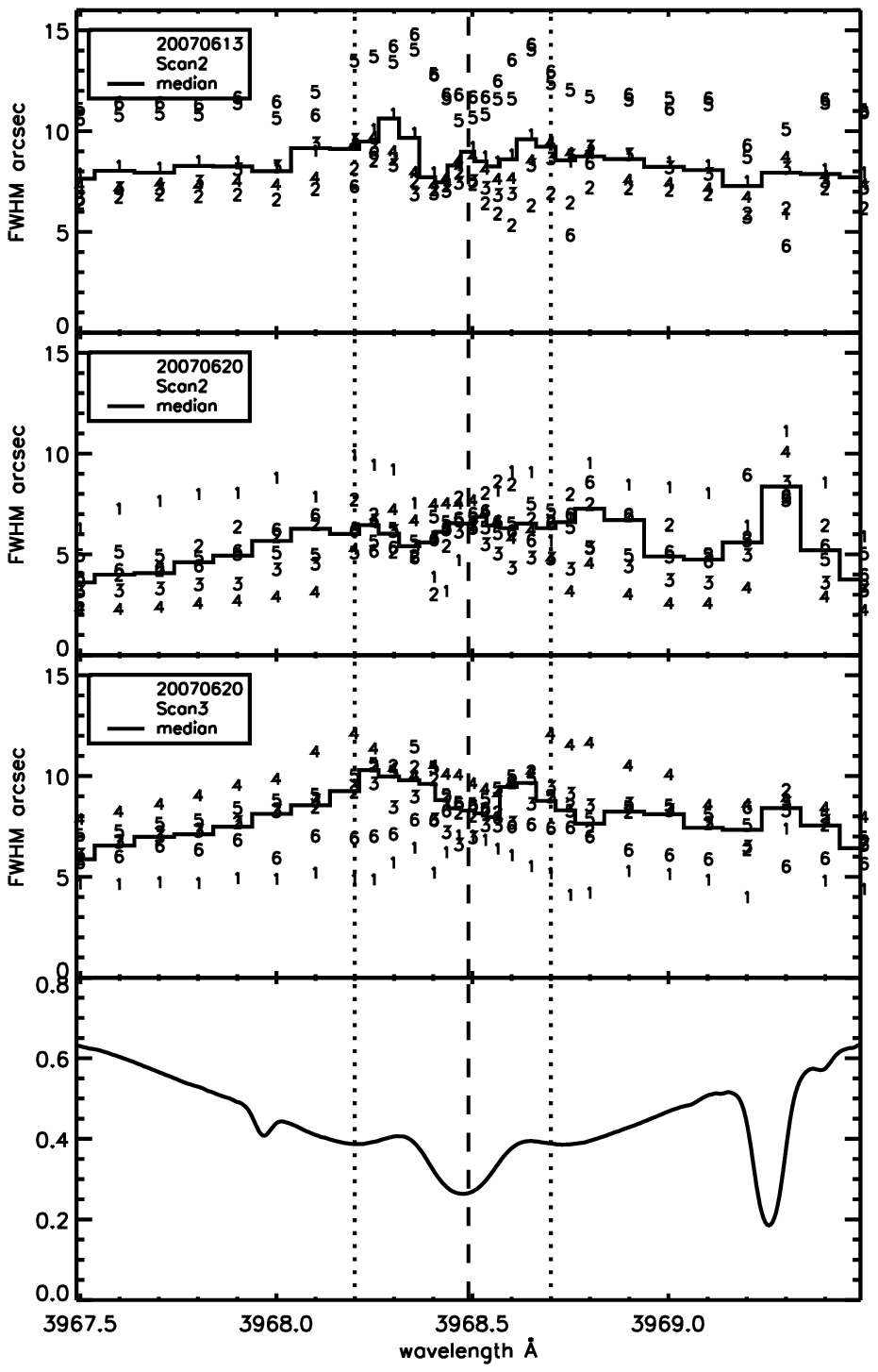}  
\caption{\label{fig:autocs} Charcteristic network widths $w_N$ plotted as a
  function of wavelength, for sub-frames of the ESG data shown in
  Figure~\pref{fig:images}
and other data (not shown) obtained on 20 June 2007.  Solid lines show
the median of the autocorrelation widths for the 6 flux
concentrations. 
The numbers correspond to data from the numbered boxes shown in earlier figures.
The lowest panel
shows the mean spectrum of the data from 13 June 2007,
where the data are normalized to the wing intensity near
3966.2 \AA.
Vertical lines
mark locations of the $H_1$ and $H_3$ minima.
}
\end{figure}
}

\newcommand\figfive{
\begin{figure}[!ht] 
\epsscale{1.0}
\plotone{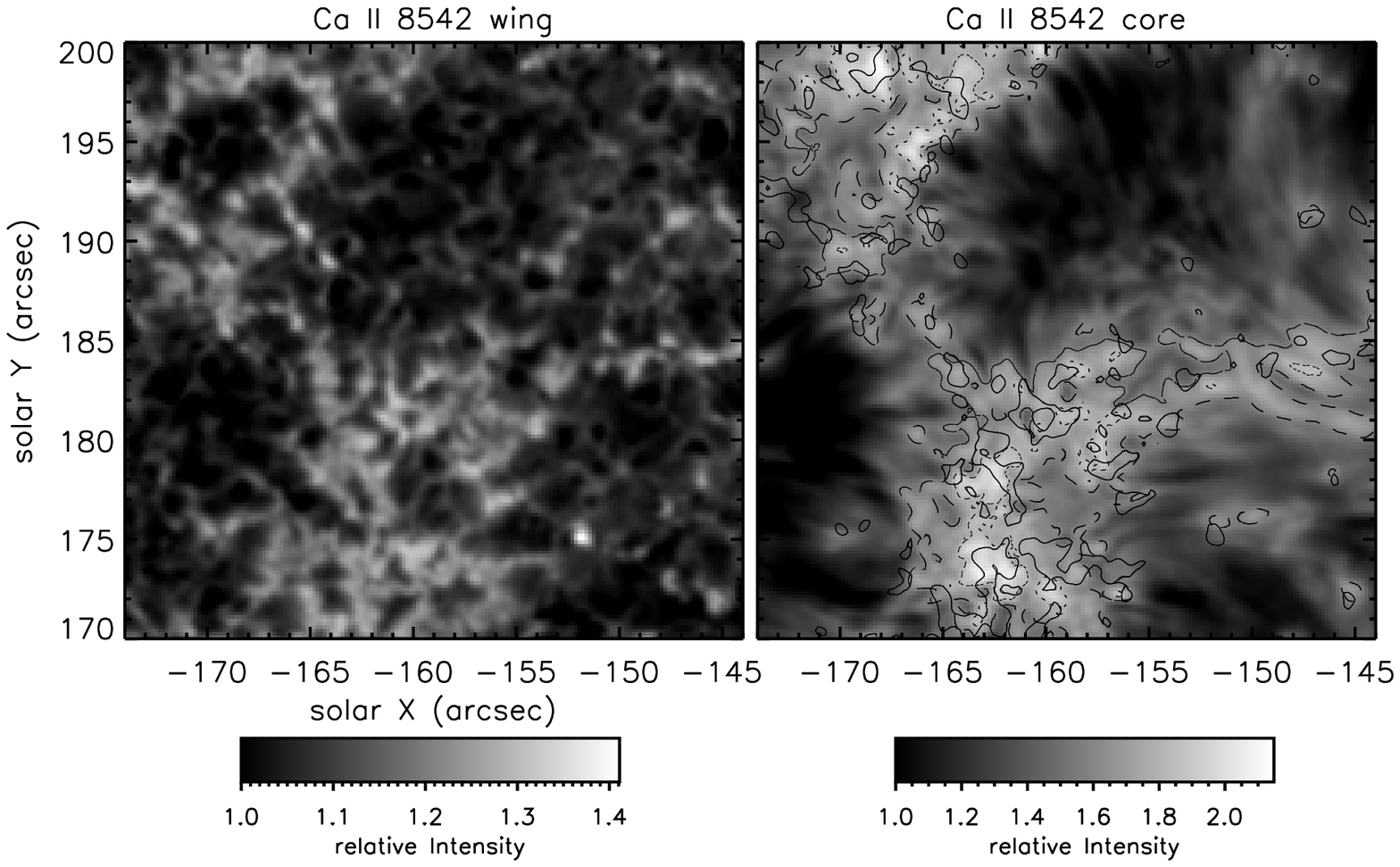}  
\caption{\label{fig:ibis_wing2core}  Wing (left) and core (right) images of
Ca II 8542 \AA  data obtained 20 May 2008 by
\protect\citet{Judge+others2010}, using IBIS.   The color tables 
vary from the lowest 5\%  of the intensity (to remove pixels from the darkest
non-magnetic features) 
to the maximum intensity
in each image.  The 50\% contour of the wing intensity (relative
wing intensity of 1.25) is over plotted in the core image (solid
lines).  The 50\% core intensity contour is shown as dashed lines, and
the 70\% contour of the core image is shown as dotted lines. 
The 50\% core contours are much broader than the 50\% wing contours,
but
the 70\% have similar areas.  Thus 
the  brightest third or so of the core emission is on small scales similar to those
of the wing image (see $X=-164,Y=173$, or $X=-167,Y=195$, for example). 
}
\end{figure}
}

\newcommand\figsix{
\begin{figure}[!ht] 
\epsscale{0.9}
\plotone{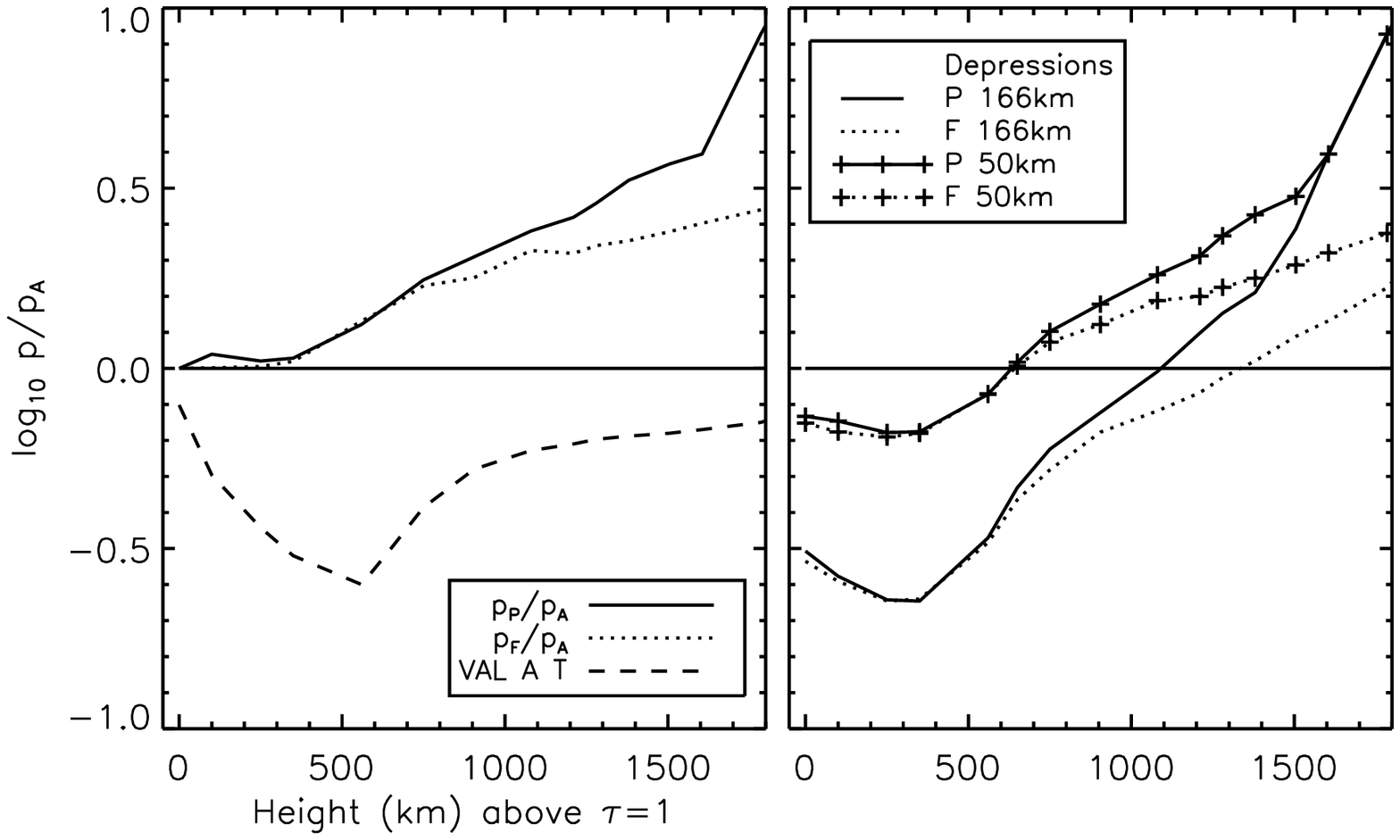}  
\caption{\label{fig:val} Pressure as a function of height in two
  ``flux tube'' semi-empirical chromospheric models (models VAL F and
  P) are plotted, relative to VAL model A.  In the left panel, the
  photosphere is assumed to be at the same physical height in each
  model. The
  form of temperature as a function of height for model ``A'' is shown
  as a dashed line in the lower half.  In the right panel, the ``tube'' atmospheres, models ``F''
  and ``P'', have been shifted downwards by 175 and 90 km.
 }
\end{figure}
}

\slugcomment{}

\begin{document}

\title{\large A chromospheric conundrum?}

\author{Philip Judge and Michael Kn\"olker}
\affil{High Altitude Observatory,
National Center for Atmospheric Research\altaffilmark{1},
P.O. Box 3000, Boulder CO~80307-3000, USA\\ \vbox{}}

\author{Wolfgang Schmidt and Oskar Steiner}
\affil{Kiepenheuer-Institut f\"ur Sonnenphysik, Sch\"oneckstr. 6, D-79104 Freiburg, Germany)}

\begin{abstract}
  We examine spectra of the Ca~II H line, obtained under good seeing
  conditions with
  the VTT Echelle Spectrograph in June of 2007, and higher
  resolution data of the Ca II 8542 \AA{} line 
from Fabry-P\'erot instruments. 
 The VTT targets were areas near disk
  center which included 
quiet Sun and some dispersed plage.  The infrared data included quiet
Sun and plage associated with small pores. 
Bright chromospheric network emission patches expand
little with wavelength from line wing to 
  line center, i.e. with increasing line opacity and height.  
  We argue that this simple observation has implications for the force and energy balance of the 
  chromosphere, since bright chromospheric network emission is
  traditionally 
  associated with enhanced local mechanical heating which increases 
  temperatures and pressures.  Simple physical
  considerations then suggest that the
  network chromosphere may not be able to reach horizontal force
  balance with its surroundings, yet the network is a long-lived
  structure.   We speculate on possible reasons for the observed behavior.
  By drawing attention to a potential conundrum,  we hope to 
  contribute to a better understanding of a long-standing unsolved problem: the 
 heating of the chromospheric network.  
\end{abstract}

\keywords{Sun: atmosphere - Sun: chromosphere - Sun: surface magnetic fields}

\section{Introduction}
\label{sec:introduction}

Over a century ago, \citet{Hale+Ellerman1904} obtained the first
spectroheliograms of the disk chromosphere.  Their work was the first
to reveal the bright ``chromospheric network'' pattern, with scales of
around 30Mm, at various wavelengths in the Ca~II $H$ and $K$ lines,
and in H$\beta$.  It is now known that the ``Ca~II network'' overlies
photospheric magnetic field concentrations clustered in supergranule
downflow lanes \citep{Simon+Leighton1963,Simon+Leighton1964,
Skumanich+Smythe+Frazier1975,Schrijver+others1989}.  

The Sun's network chromosphere emits variable UV radiation and 
sets the lower boundary conditions for the bulk of the overlying
corona.  It therefore determines and/or mediates the Sun's radiative
and particulate influence on the upper atmospheres of the earth and
planets.  Its importance in this regard stands in contrast to our 
basic understanding of it.  While 
our knowledge of the dominant physical processes in the
chromosphere is growing, the network component of the chromosphere
shows behavior which particularly continues to challenge us
\citep[see, e.g.][]{dePontieu+others2007}.  There are both observational
and theoretical reasons for this state of affairs. 
The chromosphere has a pressure
scale height of $\sim125$ km, which subtends $~0\farcs18$ at the
earth, and which is close to the
resolution of modern 
chromospheric observations. It is also structured by magnetic fields at scales down
to the limits of the instruments used.  Non-LTE radiation transport
presents a major difficulty, and the chromosphere contains highly
dynamic phenomena when observed on the smallest scales.

In this era of large numerical simulations and data of unprecedented 
angular resolution, we take a step back to 
point out a simple observational fact that, to us at
least, came as a surprise, and discuss its implications for our
current understanding of the basic physics of the chromosphere.  
Our approach is based upon observational
data and simple physical 
considerations.  While MHD models exist, we
avoid drawing heavily on these calculations 
because 
there is no credible, unique  model for
the heating of the network chromosphere, a subject of central interest
to the present work.
Our
discussion is reminiscent of an early debate concerning force balance in
sunspots \citep{Alfven1943,Cowling1976,Maltby1977,Giovanelli1982}.  

\section{Observations}

\subsection{Slit spectra of the Ca~II H line}

\citet{Rammacher+others2008} reported on data obtained near the Sun's
disk center with the Echelle Spectrograph (ESG) of the VTT on Tenerife
in June 2007.  We use a subset of these data for the Ca~II H line which they
call ``large x-y'' maps.  The Sun's surface center was
imaged onto the ESG slit, and the spectrograph dispersed the
$\approx 170\arcsec{}$ long section of the Sun's light onto a detector with
pixels every $0\farcs33$ along the slit and every 4.5 m\AA{}  in the
wavelength direction ($R=\lambda/\Delta \lambda = 900,000$).  
Each detector image was  read
into 513 spatial pixels by 976 wavelength pixels (spanning 4.4 \AA).  The
slit was rastered across the solar surface at angles tilted slightly
in the S-N direction, with
steps of $0\farcs5$, yielding a field of view of $170\arcsec
\times 120\arcsec$.  
The data were flat-field and dark corrected using standard techniques.
Table~\pref{tab:log} lists the circumstances of these observations.
Data for 13 June 2007 were of a mostly unipolar part of a 
small, decaying active region, the
other data were of quieter regions containing mixed polarities.
The seeing, measured by $r_0$, was classified as good.  
The adaptive optics system at the VTT provided good image quality and
excellent pointing stability during the raster scans.
The measured angular resolution, judged from spatial power spectra, is
typically 1.25 times the Nyqvist sampling limits of $0\farcs66$ and
$1\arcsec$, varying between 1 and 2 times these limits depending on
instantaneous seeing conditions.

Figure~\pref{fig:context} places the ESG scan for 13 June 2007 into
the context of a SOLIS synoptic magnetogram scan obtained from 14:21
UT to 14:33 UT at 6302 \AA.  In the figure the SOLIS data have been
rotated back to the time of the ESG observations, 10:54 UT.  The
corresponding line center data from the ESG are shown in
Figure~\pref{fig:core}.  In the latter figure, chromospheric fibrils
with structures down to the sampling-limited resolutions 
are
visible, confirming that the seeing was indeed good and the
instrument stable.  The boxed regions in the figure outline
concentrations of bright Ca II emission, overlying regions of
predominantly positive magnetic flux (Figure~\pref{fig:context})
associated with chromospheric network emission at various intensity
levels.  The size of these boxes was chosen as
$30\arcsec\times30\arcsec$, sufficiently large to examine the widths
of the bright Ca II network emission, but slightly smaller than the scale of
supergranulation itself.  

Figure~\pref{fig:images} shows images at selected wavelengths across the Ca~II H line for the data
obtained on 13 June 2007.  The images were constructed at each
wavelength listed for each sub-frame of 
Figure~\pref{fig:core} 
by summing over five wavelength bins (a 22.5 m\AA{}
bandpass, corresponding to a resolution of 180,000).  Images on the
short wavelength side of the line are qualitatively similar.  
In all cases,  {\em the bright chromospheric
network emission does not
expand much as the center of the Ca II H line profile is approached.  } 
The data at 1000 and 399 m\AA{} show the reverse granulation
associated with the upper photosphere, and bright network emission. 
Closer to
line center, the bright emission changes character to 
fibril-like structures familiar in the H$\alpha$ literature, but  {\em
  the emission expands very slightly, if at all.}   Instead, there is
an apparent smearing or ``filling-in'' of the granular structure
visible at wavelengths farther from line center.
 
To quantify this observation, Figure~\pref{fig:autocs} shows 
characteristic scales in the images computed from the network features centered in
the boxes shown in Figure~\pref{fig:images} and for other boxed
regions (not shown) 
for the other two 
datasets.   
Characteristic 
widths $w_N$ (FWHM) of the network features were computed as
follows. First we  subtracted
the intensity at the lowest 3\% of the intensity distribution for each
sub-image, to permit us to measure the widths of the emission peaks
above typical background levels. (The choice of 3\% is not critical).  To remove edge effects 
edges of each such sub-image were
apodized using a cosine bell function.   The 
areas of pixels $A_{50}$ which
exceed 50\% of the peak intensity were computed, and the widths were
determined 
from $\pi w_N^2/4 = A_{50}$.  The widths are therefore a characteristic FWHM of
the brightest features at the centers of the sub-images.  The results are
plotted as a function of wavelength for each sub-image 
(Figure~\pref{fig:autocs}).  

In all three datasets a trend emerges:  widths $w_N$ of 
line core images are similar to or only 
marginally larger than those of the line wings.  In 
scan 2 obtained on 13th June 2007, of plage in a small active region, 
the median data for the line cores have scales of
9-10\arcsec,  just 20\% larger than  measured at wing wavelengths. In the
other two datasets, quieter regions, the cores widths are $\lta 40\%$ larger
than the widths measured in the wings.

We conclude that Ca~II H line core images of bright network elements,
as seen at the $\lta1\arcsec$ resolution of our VTT data, 
have geometric scales only marginally ($\le 40\%$) larger than those
for the underlying wing images.  There is a hint that at wavelengths
between $H_3$ and $H_{2V}$, the widths of structures 
are narrower than the $H_2$ peaks on the red and blue side of the
line. 

\subsection{Data of higher angular resolution }

The highest resolution (0\farcs1) images of photospheric magnetic flux
concentrations reveal 
structure in the network down to the diffraction limit
\citep[e.g.][]{Berger+others2004}.   The question then arises as
to the interpretation of data with the $\sim 1\arcsec$ resolution 
data studied above.  
To study network expansion with height, a spectral resolution of
$R\gta 50,000$ is needed to resolve chromospheric line profiles
\citep{Reardon+Uitenbroek+Cauzzi2009}.
Thus, well-studied but relatively broad band Ca~II H or K line
images (e.g. from the Dutch Open telescope with $R\sim 3000$, the
Swedish Solar telescope with $R \lta 3500$, the Hinode spacecraft
$R\sim 1000$) cannot address this problem directly.  The present
generation of Fabry-Perot imaging spectroscopic instruments with
$R \gta 100,000$ and chromospheric capabilities, 
including IBIS \citep{Cavallini2006}, CRISP
\citep{Scharmer+others2008}, and
GFPI \citep{Bello-Gonzalez+Kneer2008}, can in principle shed light
on the problem addressed here. In particular, with adaptive optics and
image reconstruction techniques, such instruments can achieve far
higher angular resolution.  

Surprisingly few published studies are of direct relevance to the
problem at hand.  \citet{Leenarts+others2009} compare CRISP data for
the 8542 \AA{} line of Ca II with MHD simulations.  The bright emission
over flux concentrations is confined to subarcsecond structures in
their figure 3 showing both observations and calculations.  Care must
be taken in drawing quantitative conclusions from their work because color tables are
different between the panels in their figure.  Furthermore, the
relationship of these observations and models to conditions in
``standard'' models of the network chromosphere, discussed below, is
not clear.  

Figure~1 (panel a) and Figure 2 (panel e) of
\citet{Vecchio+others2007} show wing and core intensity images of the
8542 \AA{} line, from IBIS observations of a mixed polarity region of
quiet Sun.  Expansion of the bright network appears to be a factor of
several from wing to core.  Yet, in their data, there exist bright knots of Ca II emission at core
wavelengths, within the expanded network emission, on sub-arcsecond
scales.  Again individual color tables are different for different
panels.  To quantify the expansion of the network seen in the 
8542 \AA{} line, we have re-examined the sharpest images 
($\sim 0\farcs5$ resolution) 
of an IBIS dataset by \citet{Judge+others2010}, of small pores and
active network.  Qualitatively these images appear similar to those of
the cited work of
Vecchio and others.   The net intensities of Ca II wing
and core data are shown here in Figure~\pref{fig:ibis_wing2core},
where 
the intensity of the lowest 
5\% level in the intensity distribution has been subtracted 
to remove the darkest non-magnetic features.  
Over-plotted as contours are 50\% contours of the 
wing and core net intensity images (solid and dashed lines respectively).  The
dotted lines are 70\% core contours, shown because these 
contain a similar relative area to
the wing 50\% contours.   The 50\% intensity contours of the core are
far broader than those of the wing data. However, the network contains much
structure at the 70\%  core intensity level which lies within just a
couple of arcseconds of the underlying bright wing emission.  

These higher angular
resolution data indicate that the brightest knots of chromospheric 
network emission are
confined to within  $\sim 2\arcsec$ of the underlying photospheric
emission, but that they are often displaced signficantly from the
underlying bright photospheric magnetic features.  
Further expansion of the chromospheric magnetic field into the network cell
interiors is seen only as dimmer features associated with fibrils
expanding
into the surrounding area. 

\section{Discussion}

While significant small-scale dynamics can be observed on 
timescales down to seconds
\citep[e.g.][]{dePontieu+others2007}, the network pattern's 
lifetime is on the
order of a day. But sound waves cross a 30Mm structure in about 
an hour, a 3Mm wide patch of magnetic network boundary in a few minutes.  
The overall network structure should therefore be close
to (magneto-) static equilibrium, and   we proceed assuming this to be
the case.

\subsection{A dilemma?}

Characteristic scales of Ca~II images of bright network elements are
similar in the line core to those in the wings, yet optical depths in
the wings of the lines beyond 1\AA{} from line center are several orders of
magnitude smaller than core optical depths. One dimensional
atmospheric models \citep[e.g.][henceforth
``VAL'']{Vernazza+Avrett+Loeser1981} place the formation of the wings
and core of the H line near 0.4 and 1.9 Mm above the continuum photosphere
respectively, where the corresponding gas pressures are $\sim
5\times10^{3}$ and $\sim 4\times10^{-1}$ \dynu{}.  To satisfy force
balance, the magnetic fields must expand horizontally with height
through the chromosphere.   In the Fabry-P\'erot images, this expansion
is seen as relatively dark fibrils extending into the internetwork regions.
 {\em But why, then, does the brightest Ca~II emission
  expand far less than the magnetic field with height in the
  chromospheric network?} To address this question, we first review
conditions in the photosphere and 
corona-chromosphere transition region, afterwards discussing the 
chromosphere in this context.  This lack of observed
expansion we call ``confinement''.

\subsection{Why the photospheric network is bright and confined}

\citet{Lites+others2004} presented perhaps the clearest observations
of the bright network,  otherwise known as ``faculae'' when seen 
in photospheric features.
These observations are compatible with a well known 
physical picture \citep[as reviewed by][]{Steiner2007}.  Photospheric
magnetic fields are forced by convective flows to collect in downflow
vertices, where they exert significant pressure.  Time-averaged force
balance requires that plasma pressure inside the magnetic
concentration be lower than outside, leading to less material, less
opacity there.  The concentrations therefore allow radiation from the
``hot walls'' of the atmosphere in which they are embedded to
penetrate into them and out of the atmosphere.  This is why faculae
are bright.  

\subsection{Why the transition region network is bright and confined}
\label{subsec:tr}
 
The network seen in overlying transition region emission is also
bright, but the network pattern disappears near coronal temperatures
of $10^6$K \citep{Tousey1971,Reeves1976}.  The physical reasons that
this network is bright and confined are different from the
photospheric case.  At least two interpretations have been proposed in
the literature.

\citet{Gabriel1976} considered a magnetostatic network model 
embedded in an atmosphere with a magnetic
and a non-magnetic component.  The magnetic field, assumed unipolar,
was confined to network boundaries at the photosphere and was
space-filling high in the corona, producing the characteristic
``wine-glass'' shape of field lines.  Gabriel solved for a magnetic
field that is potential everywhere inside the wine glass, and zero
outside it, except at the current-sheet boundary between the flux concentration and the
network cell interior. (He had to solve for the location of the
boundary by requiring pressure balance across this current sheet).  He
then obtained thermal structure by imposing a balance between the
divergence of conductive flux down from the (uniform) corona and
radiation losses.  He showed that transition region emission is
confined over a several Mm wide area over 
the stem of the wine glass.  In this model, ``confinement'' of
the emission over the network occurs because energy is ducted from the
corona along field lines which are defined by the magnetic morphology,
which in turn is defined by the boundary conditions at the
photosphere, corona and the current sheet.

A second picture recognizes that such models, based on classical heat
conduction, fail to account for the brightness of lines formed in the
transition region below $10^5$K.  Some have also argued that
morphology of images of the network transition region are inconsistent
with such models \citep{Feldman1983}.  Thus
\citet{Dowdy+Rabin+Moore1986} have proposed that
network transition region is emitted from plasma confined to
relatively cool magnetic loops \citep{Antiochos+Noci1986} which close within the network
boundaries themselves to opposite magnetic polarities, and which do
not reach coronal temperatures. This picture itself has theoretical
and observational problems and remains a subject for debate
\citep{Cally+Robb1991, Judge+Centeno2008}.

The nature of the spatial confinement of bright transition region
emission is quite different in the two cases.  The cool loop picture
requires that the plasma pressures be lower than the magnetic
pressures otherwise the loop structures would tend to expand- the
confinement is due to forcing.  But, within the transition region in
Gabriel's model, the plasma pressure is irrelevant, as the magnetic
field is potential there- the confinement is caused by field-aligned conductive
energy transport.  Typical gas pressures within the transition region
network are $p\sim 0.3$ dyne~cm$^{-2}$
\citep[e.g.][]{Mariska1992,Doschek+others1998}, the corresponding
magnetic pressure requires $B\sim 10^1$G.  Yet Gabriel's solution used
$B=1$G, and he would have obtained qualitatively the same type of
confinement for a wide range of field strengths, it depending just on
the boundary distributions of magnetic field at the photosphere,
corona, and the current sheet location.  But in the case where the
plasma $\beta=p/ (B^2/8\pi)$ is $\gg 1$, any perturbation of the force
balance by plasma pressure gradients and/or gravity will force the
magnetic field to a non-force free state in which the magnetic field
can be pushed around.  Given the dynamic nature of the observed
transition region \citep{Mariska1992} and the injection of mass
from the chromosphere below \citep{dePontieu+others2007}, it may be that Gabriel's
potential field calculation is a singular case which may not occur in
reality.  Taken together, it seems that confinement both by
field-aligned energy transport and by force balance ($\beta \le 1)$
may be required to explain the observed properties of transition region plasmas.

\subsection{Why the chromospheric network is bright}

Traditionally, the chromosphere is believed to be bright because of
mechanical heating \citep[e.g.][]{Osterbrock1961}.  The required
dissipated energy flux density, derived
from the need to account for radiation losses computed from 1D
semi-empirical models, is on the order of $10^7$ \flxu{}
\citep{Avrett1981, Anderson+Athay1989}, some four orders of magnitude
smaller than the photospheric radiative flux density.  Such
calculations remain the best way to estimate the energy requirements,
since they cannot be derived from observations alone for varied
reasons.  Yet the calculations, based upon observations which do not
resolve structure below granular scales, contain little of the physics
of magnetic concentrations.  The photospheric parts of these models
are almost identical, containing no multi-dimensional transfer such as
the hot wall effect requires.  As a result, any hot wall radiation
re-radiated by the chromosphere is entirely ascribed to mechanical
heating.  We return to this issue in section \pref{sec:radheat},
But first we argue that a difficulty arises when trying to understand why
an entirely mechanically heated chromospheric network is {\em
  confined}.

\subsection{Why is the bright chromospheric network confined?}

While credible physical models of the chromosphere outside of magnetic
concentrations are available \citep{Carlsson+Stein1995,
Carlsson+Stein1997}, none is available for the magnetized regions.
We must deal instead with the semi-empirical models of network that
are available, such as from VAL, recognizing their limitations along the
way.

Chromospheric heating results in larger temperatures, and gas
pressures, at a given height.  Semi-empirical models of bright
chromospheric features, such as those over 
network boundaries, are always accompanied by
higher gas pressures and energy densities.  
But this is precisely the place where magnetic
pressures are also higher.  Herein lies a possible problem.  Taken at
face value, in particular the run of gas pressures with height, 
the semi-empirical models cannot be in horizontal force
balance, as the network boundary model would be expected to expand
horizontally until pressure exerted by neighboring magnetic fields
and/or the magnetic tension force can balance the excess total
pressure!  

This dilemma is resolved in part through explicit recognition of 2D
effects and the fact that the height scales of such models are defined
with respect to the radial 5000 \AA{} continuum optical depth unity
surface ($\tau_{5000}=1$).  The hydrostatic equation used in the
models yields $p$ as a function of height $z$ only to within a
constant integration factor.  In 2D, horizontal force balance requires
gas pressures within photospheric flux tubes to be smaller at
each height than outside it.  The opacity is reduced, and the
$\tau_{5000}=1$ surface is thus shifted downwards (the ``Wilson
depression'').  On this basis \citet{Solanki+Steiner1990} and 
 \citet{Solanki+Steiner+Uitenbroeck1991} built ``1.5D'' models of flux
tubes, endowed with thermal conditions of hotter and brighter models, e.g. VAL
model F, embedded in
cooler models.  The entire flux-tube atmosphere (not just the photosphere)
was moved downwards by $200$ to 500 km, to satisfy horizontal 
force balance within the photosphere, 
depending on the combination
of thermal structure and field strength used. The authors examined
models with field strengths between 1500 and 1630G. Higher up, these authors iterated
the magnetostatic equation to equilibrium, keeping the thermal
structure within the two components fixed, both being  a function of height
only.
Above a certain height (the ``flux merging''
height), although the plasma by itself cannot reach horizontal force
balance (see Figure \pref{fig:val}), in the models of Solanki and
colleagues the magnetic field bumps into its neighbors, thereby taking
up the net horizontal pressure gradient.  In these 1.5D models,
$\tau=1$ surfaces of network chromospheric features are geometrically
lower than those of the non-magnetic surroundings.  The emission is
effectively confined and our dilemma appears resolved, at least qualitatively.

However, consideration of the physics underlying these
calculations leads us to conclude that the ability of 
such 1.5D models to achieve magnetostatic equilibrium is
remarkable.  Wilson depressions of the magnetized
photosphere are determined by the force balance only at {\em
  photospheric} heights.  But the variation of {\em chromospheric}
plasma pressure with height is determined largely by the unknown
process(es) of chromospheric heating which occur many scale heights
above the photosphere.  Further, thermal structures 
in these semi-empirical models are in large part based upon
observations, with no explicit treatment of force balance except
through the constraint of vertical (1D) hydrostatic equilibrium.
There is in fact {\em no prior reason to expect that
  the physics of {\em chromospheric} heating will accommodate the
  conditions needed for 3D magnetostatic equilibrium there, simply by
  dropping the network atmosphere by the amount needed to bring the
  {\em photosphere} into horizontal pressure balance.}  
We raise additional concerns. 

\begin{itemize}
\item{} $B=1500$G is near the high end of a broad distribution of
  observed photospheric network field strengths
\citep{Berger+others2004} and in MHD models
  \citep{Schaffenberger+others2005, Wedemeyer-Bohm+others2007}.  Lower
  values of $B$, keeping other parameters constant, means larger
  values of $\beta^\ast$ (the asterisk refers to the value of $\beta$
  measured within the flux concentration, at the geometric height
  of the non magnetic photosphere).  For a pressure scale height $h$,
  and given a value of plasma $\beta^\ast < 1$, the depression
  $\delta$ of the photosphere required to bring it to photospheric
  force balance is $\delta \approx h \, ln \left ( 1+{\beta^\ast}^{-1}
  \right )$.  \citet{Solanki+Steiner+Uitenbroek1991} present
  calculations where $0.02 \le \beta^\ast \le 0.3$. The photospheric
  gas pressure is $1.2\times10^5$ \dynu, so for $B=1500,1000,500$G we
  find $\beta^\ast=0.34,2.0,11$, and with $h=120$ km,
  $\delta=166,49,10$ km respectively.  Even when $\delta\sim166$ km
  the gas pressures alone in models F and P exceed those of model A at
  chromospheric heights above 1.2 Mm (Figure~\pref{fig:val}). When 
  $\delta < 50$ km essentially the entire chromospheric gas pressures 
of models P and F exceed those of model A. 

\item{} At the very edge of photospheric faculae, there is no
  neighboring flux tube to balance excess pressure, a situation yet
  more dramatic when the embedding atmosphere has no chromospheric
  temperature rise such as is the case in the dynamic models
  \citep{Carlsson+Stein1995}.  Fig.~2 of \citet{Solanki+Steiner1990}
  implies that the gas pressure surpasses the outside pressure already
  at 800 km and 850 km for B=1300 and 1500 G, respectively.
\end{itemize}

There appears to be no easy way to make horizontal forces balance
across bright network boundaries in the chromosphere in such models.
At the very least we must understand why the chromosphere produces the
brightest emission only directly over the photospheric magnetic
concentrations.

\subsection{What is the chromospheric plasma $\beta$ in network boundaries?}
\label{subsec:beta}

In both the chromosphere and transition region, the plasma $\beta$ 
is a critical parameter, since as in plasma devices, confinement by
magnetic forces requires 
$\beta < 1$.  We expect $\beta$ to vary both 
along and between  field lines through the chromosphere, as the flux
tubes are in no sense ``thin''.   The atmospheric stratification 
is strong (scale heights $h$ are $\sim120$ km in a chromosphere of thickness
$\sim 1500$ km), so that we expect $\beta$ generally
to decline with height. 
Upper limits to $\beta$ extending though the chromosphere can be
estimated using simply the net magnetic flux density averaged over a
supergranular cell, assuming it is homogeneous.  
In quiet regions the mean field strength will be
typically zero but with fluctuations of a few tens of Mx cm$^{-2}$.
In more active network and plages this will increase to $\gta 100$ Mx
cm$^{-2}$, in sunspot umbrae (for comparison) it will be $\gta 2000$
Mx cm$^{-2}$. Applying these estimates to VAL's model F, we expect
$\beta=1$ in quiet, plage and umbral regions to occur below heights of
1.2, 0.65 and -0.05 Mm respectively.

Direct observational determinations of $\beta$ are rare because
magnetic measurements using chromospheric features is difficult.  Most
such work has been concerned with sunspots
\citep{Metcalf+others1995,Socas-Navarro2005}, but network fields were
studied by \citet{Pietarila+others2007a,Pietarila+others2007b}.  The
latter were made in plages associated with a decaying active region.
\citet{Pietarila+others2007b} made inversions for a region of
predominantly negative polarity with a net mean flux density of order
50 Mx~cm$^{-2}$.  (This value was estimated by eye from Figure~2 of
\citealp{Pietarila+others2007a}.)  The longitudinal component of the
field strength could be determined only near $\tau_{5000} =0.1$ and
$10^{-5}$.  Using the range of values of field strengths from their fig. 12 (network
element), and assuming thermal parameters from VAL's model F, the
plasma $\beta$ values are probably between 5 and 100 (at 0.15 Mm height)
and 0.1 (at 1.1 Mm). The $\beta=1$ level probably lies 
near 0.5 Mm.

In truly quiet regions no measurements exist, but likely values of the
plasma $\beta$ can also be examined using ``realistic''
simulations\footnote{We have deliberately avoided use of 
  simulations for the interpretation of chromospheric brightness,
  because brightness depends exponentially on the temperature and the
  heating mechanism(s) are not known.  Our use here is also
  questionable, but estimates of the plasma
$\beta$ from simulations are arguably more reliable, at least for the
first few scale heights of the chromosphere.}. 
\citet{Schaffenberger+others2005,Schaffenberger+others2006} have made simulations
of magnetconvection extending 1.4 Mm above the photosphere, for an
average field of 10 Mx cm$^{-2}$.  In this case, there is not enough
flux for the granulation to gather up and form the most intense 1.5 kG
photospheric flux tubes- their field strengths are $\le 1$ kG.  The
computed $\beta=1$ surface lies near 1.2 Mm above the photosphere,
i.e. well within the chromosphere, yet in their calculations no
explicit or significant chromospheric heating was included to enhance
chromospheric temperatures and pressures.  The existence of field
strengths mostly at or below 1 kG in the photospheric network
\citep{Berger+others2004}
implies that $\beta \ge1$ in the lower parts of the chromosphere
(\citealp{Hasan+vanBallegooijen2008}).

We conclude that the lower 0.5-1 Mm or so of the network chromosphere is
probably in a high $\beta$ regime, depending on local conditions.  

\subsection{Mechanisms for confining chromospheric network emission}

\subsubsection{Radiative energy transport}

\label{sec:radheat}
It is almost universally believed that the chromosphere is bright
because of non-radiative, or mechanical, heating. However, Uitenbroek
(private communication, 2009) has pointed out that Ca II H emission
reversals can occur when the photospheric ``hot wall'' radiation
raises the source functions and brightness of the Ca II wavelengths
which are normally considered chromospheric.  Such bright emission
from the chromosphere is indeed expected to be confined to within a
volume where the hot wall radiation can influence the source
functions, the ``thermalization volume''.  2D or 3D radiative
equilibrium calculations are needed not only to relax the constraint
of LTE, but also to take proper care of energy transport in spectral
lines.  The treatments made so far including opacity distribution
functions and multi-group methods \citep{Steiner1990,Skartlien2000}
all adopt the coherent scattering approximation, thereby artificially
reducing the range of influence of energy transport in spectral lines.
The thermalization lengths are too low, the lambda- and other
operators for lines are spatially too narrow.  It remains to be seen
if a more physical treatment of lines leads to significant differences
with these calculations.  Interestingly, 
\citet{Bruls+VonderLuhe2001}
found only small intensity enhancements in the $H_{2R}$
peak and $H_{1R}$ minimum, 
from a 2-D hot wall nLTE calculation again using semi-empirical 
models, suggesting that this effect may be small.  But no
self-consistent radiative equilibrium model has yet been performed.

However, transport of hot wall radiation cannot 
fully resolve the present dilemma, because it cannot account for
emission lines requiring electron temperatures in excess of the
temperature of the hot walls, such as UV and EUV lines.  Nor can it
explain
the properties of bright knots of chromospheric emission in the line
cores in data such as those shown in Figure~\pref{fig:ibis_wing2core}.  
Also, the
opacity may simply be just too large to permit transport of much
energy from the hot walls into the body of the chromosphere. 
Nevertheless
we note that essentially all mechanical heating requirements of the
chromosphere are based on 1D semi-empirical models which do not
include hot wall radiation.  These requirements may therefore have
been uniformly over-estimated, and deserve attention.

\subsubsection{Stresses on network boundaries}

Network cell interiors are not static, field-free structures. The
possibility arises then that work might be done on the supergranular
network by the horizontal components of waves, flows and by magnetic
stresses associated with the cell interior chromosphere.  Outside of
shock waves and the eye-catching 
type-II spicules \citep{dePontieu+others2007} 
which are insignificant components of the bulk chromosphere, being 
phenomena of the more tenuous upper chromosphere 
\citep{Judge+Carlsson2010}, 
spectral line shifts and proper motions indicate sub-sonic motions. Thus
flows and waves can make only small contributions to the work done on
supergranule boundaries compared with thermal pressure gradients.

Magnetic fields in the cell interiors have received increasing
attention in recent years \citep[e.g.,][]{Lites+others2007a}. 
Such fields may perhaps provide additional stress
at the network boundaries, but they would need to compete with the
strong network boundary fields to contain the excess pressure implied
above.  There is no evidence that these fields are strong 
enough to maintain a force at the edges of network boundaries
sufficient to contain significant plasma over-pressure, if present. 

\subsubsection{Pinch effect}

Simple flux ribbons and untwisted flux tube-models 
have no degree of freedom which might allow
enhanced plasma pressures when $\beta \gta 1$.  However, if the bright
Ca II network emission is associated with 
strongly twisted magnetic fields, 
the twist 
can ``pinch'' the plasma in network boundaries, potentially 
enhancing the pressures where magnetic fields are strong.  It seems
unlikely that the large body of data supports this picture, given that
much of the network appears to be associated with flux sheets
in granular downflow lanes  \citep[e.g.][]{Berger+others2004}, but 
the consequences of such a picture are of potential interest. 

Many theoretical twisted flux tube models have been made, beginning
with \citet{Parker1974,Parker1977}.  Parker studied the analytical properties of
force-free tubes which have slowly varying radii $r$ with
distance along the tube $z$, i.e.  $\partial r/\partial z \ll 1$.  His
essential finding is that the degree of twist increases with the
tube's radius, the field becoming entirely azimuthal when the radius
exceeds a critical level.  The increasing twist results from the
conservation of magnetic flux and torque (equivalently, electric
current) along the tube.  As a consequence, expanding tubes suffer
increasing twist which eventually overcomes the pressure gradient
force. Parker invoked this ``pinch'' or ``buckling'' effect 
as a way to release magnetic free energy in the much larger scale context of 
emerging solar active
regions. Non-linear numerical calculations were made, for example, by
\citet{Steiner+Pneuman+Stenflo1986}, confirming Parker's 
picture.  By examining conditions high in the atmosphere where flux
tube merging has occurred, these authors derived an upper limit
to the twist at the tube's base, $B_\phi/B_z < \sqrt{f}$, where $f$ is
the area filling factor of the tube at its base. This limit was
obtained in the low-$\beta$ limit when 
the tension force $\sim B_\phi^2/4\pi r $ exceeds magnetic
pressure gradients $\frac{1}{8\pi}\pder{B^2}{r}$ within the tube.  It is precisely
near this limit that the pinch may therefore be expected to compress
the plasma in a higher $\beta$ regime.  
In the quiet Sun,
$f\lta 0.01$ or so, so that $B_\phi$ can at most be 10\% of $B_z$ in
the photosphere.
Beyond this twist, the merged fields in the upper part of the tube
become unstable to the pinch effect. The evolution after pinch onset is not
known.

The plasma pressure differences in the VAL cell and boundary models A
and F are on the order of the pressures themselves, so that the twist
must be near this limiting case if it is to account for the inferred
thermal properties.  In this case then the field becomes predominantly
azimuthal.  When $f=0.01$, this requires that the tube expands
radially tenfold over the value at the base. For a flux tube with base 
radius of 0.1 Mm the chromospheric twisted tube radius would become
1Mm, a value not inconsistent with the observed widths of the
supergranular network.  The dynamic consequences of such a
situation are speculated upon below.

A recent letter reports Mm-scale swirling chromospheric motions above network
elements in coronal holes
\citep{Wedemeyer-Bohm+Rouppe-van-der-Voort2009}, apparently driven by 
random granular motions below.  Such motions, visible in the
relatively uncluttered magnetic field in coronal holes, 
are probably
related to this scenario.  It is notable that  there is no obvious relationship
of the observed chromospheric brightness to these swirling motions. 

\subsubsection{Steady-state dynamics vs. static equilibrium}

A different resolution of our dilemma may lie in the 
possibility that the system is never close to equilibrium,
but appears so when seen with existing instruments.  By 
analogy, one might consider a cloud pattern which appears stationary
when seen from a large distance, but closer observation  reveals 
a non-steady dynamic evolution of individual clouds. 

The intensity of chromospheric features like Ca II H scales exponentially with
temperature, so that positive
fluctuations in space and/or time in $T$ might produce emission
sufficient to account for the observed cell/ boundary intensity
differences, while maintaining approximately magnetostatic balance in
horizontal planes.  This kind of picture is valid, at least in part,
in the cell interior regions
\citep{Carlsson+Stein1995,Carlsson+Stein1997}. Acoustic gravity waves
there propagate upwards and shock near 1 Mm above the photosphere to produce
occasional bright bursts of chromospheric emission on top of a weaker,
unresolved background emission.  The time-averaged emission in their
computations of \ion{Ca}{2} lines, where $h\nu/kT \gg 1$, arises from
a plasma in which the average temperature is far lower than would be
required to produce the emission in a static model.  In this case the
average pressure is also less than a static model would require.  The
difference between the network boundary and cell interior problems is
that there is much observational support for the waves in the latter
case, but our knowledge of the mechanism(s) of network boundary
heating and dynamics is not at all clear \citep[see,
for example][]{Lites+Rutten+Kalkofen1993a,Judge2006}.

Suppose, as in the cell interior
case, that energy is released in the network boundary chromosphere
intermittently but on time scales less than the boundary wave
crossing time of $\tau_w \sim w/c_s\sim5$ minutes.  Here $w\sim 3$
Mm is characteristic scale of 
the network boundary thickness,  $c_s$ is the sound speed, and  we assume
$\beta \gta 1$. 
The time scale for reaching ionization
equilibrium is $\sim 40$ seconds, being determined by
the long times needed 
for hydrogen to recombine.  Energy released as heat can be stored as
latent heat of ionization on time scales longer than this.  
The radiative relaxation time
$\tau_{rr}$ of chromospheric plasma is
$$
\tau_{rr} \sim \frac{1}{\gamma-1} \frac{nkT}{\Phi} ,
$$ where $\Phi$ is the radiative energy loss rate per unit volume.
When perturbations occur faster than the ionization time scale 
we can use $\gamma=5/3$, since no change of internal state occurs.
From figure 4 of \citet{Anderson+Athay1989} we find $\Phi \approx 300{m}$
erg~cm$^{-3}$~sec$^{-1}$ for heights $>1$Mm (column mass 
$m<10^{-3}$ g~cm$^{-2}$).   This value applies to 
average quiet Sun conditions- $\Phi$
for the hotter network will be larger.
Hydrostatic equilibrium gives 
$p=nkT=mg$, with $g$ the gravitational acceleration,  and the radiative cooling time becomes 
$$
\tau_{rr} \lta \frac{3}{2} \frac{mg}{300m} \sim 140 \ \ \ {\rm sec,} 
$$ which is independent of $m$ and smaller than $\tau_{w}$.  ($\tau_{rr}$ increases
rapidly in deeper layers owing to radiative transfer effects.)
Thus, chromospheric plasma can in principle store energy
as latent heat of ionization and radiate it faster than sound waves can
communicate any overpressure to the network cell interiors.  This
state of affairs is precisely what is needed to sustain a bright
non-equilibrium network chromosphere heated intermittently on time
scales shorter than $\tau_w$, while at the same time
avoiding a sustained large horizontal pressure imbalance between the
network boundary and the cell interior.

This suggestion might have support observationally, as the network
chromosphere is the site of significant small scale
``activity''. Spicules originate from these regions
\citep{Beckers1972}.  Time series spectral observations, including the
chromospheric \ion{Ca}{2} $H$ line, obtained by
\citet{Lites+Rutten+Kalkofen1993a}, show marked differences between
network cell boundaries and interiors.  Cell boundaries show
variability on time scales of 5 minutes and longer, which appear to show
features propagating slowly away from the peak of the network
emission, with apparent speeds of a few km/s 
($\lta 1$ to 2 Mm in 5 minutes). These features repeat perhaps every
8-10 minutes.  There is as yet no accepted physical model for this network
behavior. 

Recent MHD simulations extending from the convection zone to 1400 km
above the photosphere also lend credibility to this
possibility\footnote{Again we draw on the simulations to examine
only the nature of the dynamics of the convectively driven 
magnetic field in the overlying atmosphere and avoid drawing on
properties strongly coupled to the energy equation.}. \citet{Schaffenberger+others2005,
Wedemeyer-Bohm+others2007} have found that the magnetized part of the
chromosphere behaves quite differently from the underlying
photosphere, albeit for average magnetic flux densities near
10 Mx~cm$^{-2}$, weaker perhaps than present in the data examined
here.  Nevertheless, 
 while the calculations support the general idea of quasi-static 
``canopy'' fields, the magneto-fluid over the photospheric flux
concentrations is highly variable in space and time, involving a
``continuous reshuffling of magnetic flux on a time scale of less than
1 minute''.  Much of the variability is caused by supersonic flows and
shocks, which compress the magnetic fields, especially near the $\beta=1$
surfaces.

Further theoretical support may be found in magneto-acoustic shocks
which may develop
preferentially in the cores of the magnetic network, because there the
magnetic field expansion is relatively weak.  Field-aligned slow mode amplitudes
can therefore grow faster in the cores than elsewhere, and will shock
lower in the core of the stratified flux tube atmosphere. Indeed 
\citet{Fawzy+others1998} found the tube expansion profiles to be of
critical importance for longitudinal shock-wave heating, and
\citet{Khomenko+others2008} found different properties in longitudinal
non-linear wave propagation as a function of axial distance in two-dimensional
flux-tube models. 

In essence, this proposal allows the chromosphere to radiate more
without an accompanying increase in the gas pressure and energy
density, something which cannot be achieved in static models.  We have
loosened the (overly-?) restrictive link between radiative energy flux
and gas pressure and energy density.

\subsubsection{Emission from between flux tubes?}

High resolution data show that photospheric magnetic fields
under network boundaries are collections of granular driven field
concentrations which, when observed at lower angular resolution,
become organized into the familiar chromospheric network.  The
possibility arises that the ``network boundary chromosphere'' is then
filled with the upward extension of this intermittent field, with a
mixture of field strengths and directions which become more uniform
with increasing height.  It is possible that the UV emission from the
network arises predominantly from material {\em between} these
magnetic flux sheets.  In this way the pressures of strongly
radiating material will be larger than the material embedded within
the flux tubes (B.C. Low, private communication 2007).  If this proves
to be the case, the chromospheric magnetic polarization signatures
should be detectably smaller than photospheric extrapolations would
indicate.  Magnetic signatures are present in chromospheric lines 
so that not all of the emission can be from field-free plasma
\citep{Giovanelli1980, Pietarila+others2007b}, but such a quantitative
comparison remains to be done.  
Note that such a picture does not discount magnetic
heating mechanisms, because the kinetic energy in, e.g., transverse wave motions
of the flux tubes will move  field-free neighboring
fluid,  leading perhaps to dissipative compressive waves, for example. 

\section{Conclusions and further speculations}

Natural explanations for the local 
confinement of the bright photospheric and transition region plasmas  to 
the network boundaries  have existed for many years.
However,  we have identified a
possible problem in trying to explain why the chromospheric network
emission should also be confined locally, assuming, as seems
unavoidable, that increased brightness is associated with increased
dissipation of mechanical energy and hence increased temperatures and
pressures. 
  Vertical
forces may not be sufficient to compress the chromospheric network
plasma sufficiently to account for the intense radiation originating
there, while at the same time maintaining horizontal pressure balance.
The atmosphere has time to equilibrate pressures from the
network boundary to the cell interiors (several tens of minutes)
compared with the life time of the supergranular structures (30 hours
or so), so it is not obvious that magnetostatic equilibrium is a poor
approximation on supergranular scales.  

Radiation-MHD calculations should ultimately resolve the conundrum posed, firmly
constrained by 
simultaneous spectropolarimetry of the photosphere/chromosphere.  
Such observations will 
help clarify issues such as the plasma-$\beta$ state, if field twist
is important, or if bright network chromospheric emission arises
almost entirely from material preferentially {\em between} regions of
strong field.
We can speculate that the resolution of the problem may also
provide natural explanations of the time-dependent network boundary
chromosphere and spicules. If the network heating is intermittent,
then matter will be temporarily over-pressured and will be forced
vertically along field lines and expand the field horizontally.
Should the pinch effect be important, its non-linear development may
explain both chromospheric heating and the ejection of spicules, some
of which have significant twisting motions as seen in Hinode
 \citep{Suematsu+others2008} and later data
\citep{Wedemeyer-Bohm+Rouppe-van-der-Voort2009}.  The system may be
self-limiting in that more pinch implies more confinement and
adiabatic heating, which leads to higher pressures and less
confinement (spicule emission?), and so on.  This picture is
reminiscent of, but different to, the proposition by
\citet{Athay2000,Athay2002} that an ionization instability, caused by
the preferential heating of ions in the upper chromosphere, amplifies
variable heating rates which in turn lead to the waxing and waning of
spicules.  The dynamics we envisage might also be related to that seen
in 1.5D numerical simulations of randomly driven Alfv\'en waves in
flux tubes by \citet{Kudoh+Shibata1999}.

It may be that some Ca~II emission is simply 
re-radiated ``hot-wall'' radiation from the
photosphere. It would be interesting to see 2D radiative equilibrium
calculations of flux tube atmospheres 
instead of semi-empirical thermal structure.  
This point deserves attention since it is unclear how much of what
is traditionally attributed to in-situ chromospheric heating, could instead be
simply re-radiated photospheric radiation.    

Whatever the outcome, we hope that this discussion will lead
to a better understanding of the essential ingredients of the physics
of the magnetic network chromosphere, a long standing
unsolved problem.  It represents by far the biggest ``heating
problem'' in solar physics, needing at least an order of magnitude
more energy to sustain it than the overlying corona.

\acknowledgments PJ thanks Yuhong Fan, Boon Chye Low, Mattias Rempel
and Han Uitenbroek for many interesting discussions.  The VTT is
operated by the Kiepenheuer Institut f\"ur Sonnenphysik at the Spanish 
Observatorio del Teide of the Instituto de Astrof\'isica de Canarias. 
IBIS was constructed by INAF/OAA with contributions 
from the University of Florence, the University of Rome, MIUR, and 
MAE, and is operated with support of the National Solar Observatory. The NSO 
is operated by the Association of Universities for Research in Astronomy, Inc., 
under cooperative agreement with the National Science Foundation.
We are indebted to DST observers Mike Bradford, 
Joe Elrod and Doug Gilliam.   We thank the anonymous referee for very useful comments.  

\def\aspcs{{ASP Conf.\ Ser.}}

\protect\begin{deluxetable}{lllll}
\tablecaption{Log of Ca~II H line observations with the VTT ESG \label{tab:log}}
\tablehead{Date/time & scan & pointing & raster & seeing \\
 UT  & number &  & parameters & $r_0$ cm }
\startdata
13 June 2007 10:52-10:56 & 2 & E2.4$^\circ$, S4.0$^\circ$ & 241 steps,
1 sec/step & 7-13\\
20 June 2007 8:25-8:31 & 2 & E0.0$^\circ$, N1.7$^\circ$ & 240 steps,
1.5 sec/step & 9-12\\
20 June 2007 8:39-8:45 & 3 & W19.3$^\circ$, S6.0$^\circ$  & 240 steps,
1.5 sec/step & 9-13\\
\enddata 
\end{deluxetable}

\figone

\figtwo

\figthree

\figfour

\figfive

\figsix
\end{document}